 \documentclass[final,1p,times]{elsarticle}

\usepackage{amssymb}
\usepackage{amsmath}
\usepackage{bm}
\usepackage{chemformula}
\usepackage{braket}
\usepackage{siunitx}
\usepackage{adjustbox}
\usepackage{tabularx}
\usepackage{makecell}
\usepackage[flushleft]{threeparttable}
\usepackage[normalem]{ulem}

\journal{Reviews in Physics}

\begin{document}

\begin{frontmatter}



\title{Nonlinear optical metasurfaces empowered by bound-states in the continuum}

\author{Ji Tong Wang\fnref{label1}}
\fntext[label1]{Present address: Emergent Photonics Research Centre, Department of Physics, Loughborough University, Loughborough LE11 3TU, United Kingdom.}
\ead{jitong.wang@ucl.ac.uk}
\author{Nicolae C. Panoiu \corref{cor1}} 
\ead{n.panoiu@ucl.ac.uk}
\address{Department of Electronic and Electrical Engineering, University College London, London WC1E 7JE, United Kingdom}
\cortext[cor1]{Corresponding author}

\begin{abstract}
Optical bound-states in the continuum (BICs) have greatly enriched the field of nonlinear optics with novel ways to control and manipulate light-matter interaction at the nanoscale. This has been made possible by their unique physical properties, including effective confinement of light, non-trivial topological features, and robustness upon the propagation of the optical field both in the real and momentum space. Regarding the exploration of nonlinear optical response in various photonic nanostructures supporting BICs, particular attention has been paid to optical metasurfaces, chiefly due to their ability to control the light flow at subwavelength scale, design and fabrication flexibility, and convenient phase-matching conditions. In this review, we outline and discuss recent advances in metasurface-based frequency conversion processes utilizing the versatile physics of BICs, with a particular emphasis on the main physics background pertaining to nonlinear optical phenomena and optics of BICs, as well as state-of-the-art functionalities enabled by BIC-driven nonlinear metasurfaces. These applications include harmonic generation, harmonic chiroptical effects, generation of complex quantum states, and broadband terahertz generation. In addition, several emerging research fields and the existing challenges of photonic nanodevices relying on BICs are discussed.
\end{abstract}

\begin{keyword}
Bound-states in the continuum \sep Nonlinear optical metasurfaces \sep Nonlinear optics \sep Light-matter interaction \sep Harmonic generation \sep Frequency mixing



\end{keyword}

\end{frontmatter}



\section{Introduction}
\label{sec1}

Metasurface-based planar nonlinear optics has recently emerged as a particularly active field of research \cite{Mino2015,Li2017,Krasnok2017,Panoiu2018,Sain2019,Pert2020,Vabi2023}, which integrates key optical properties of periodically patterned nanostructures with the principles governing light-matter interaction in the nonlinear optical regime. Consisting of planar distributions of artificial nanoresonators with subwavelength spatial features, two-dimensional (2D) metasurfaces provide an appealing potential for achieving tunable, precise, and flexible control of light-matter interaction at the nanoscale. They exhibit notable advantages when compared to their three-dimensional (3D) counterparts (metamaterials) \cite{Zhelu2012,STau2017,Kadic2019,Xiao2020}, including smaller number of design parameters and the availability of more powerful and versatile nanofabrication techniques.

Importantly, the ability of metasurfaces to induce significant enhancement and confinement of the local optical field makes them an ideal platform for nonlinear optics, forming the foundations for the research topic called nonlinear optical metasurfaces. In addition, the relaxation of phase-matching conditions, a restricting constraint placed upon efficient frequency conversion processes pertaining to propagating optical waves in bulky nonlinear optical media, allows one to apply metasurfaces to a broad range of nonlinear optical applications. Thanks to their unique physical properties and the availability of mature fabrication capabilities, we have witnessed in recent years the use of nonlinear metasurfaces in successful demonstration of various nonlinear optical interactions, including second-harmonic generation (SHG) \cite{Fan2006A,Fan2006B,Valev2014,Kruk2015,Liu2016,Chen2016,Valev2018,Vabi2018,Fedo2020,Mann2023}, third-harmonic generation (THG) \cite{Smir2019,Gando2021,Okhl2021,Zheng2023}, high-harmonic generation (HHG) \cite{Liu2018,Shch2021,Zogra2022}, spontaneous parametric down-conversion (SPDC) \cite{San2021,San2022}, optical rectification (OR) \cite{Hale2022,Hu2022,Peter2024}, four-wave mixing (FWM) \cite{Liu2018FWM,You2020,More2024}, and harmonic chiroptical effects \cite{Hu2019,Shi2022,Koshelev2023ACS,Tonkaev2024}. Spurred by these fast-paced developments at fundamental science level, nonlinear optical metasurfaces have facilitated rapid technological advancements in key areas of science and engineering, including information processing, sensing, quantum optics, and ultrafast optics \cite{Li2017,Krasnok2017,Huang2020,Hu2021}.

Early studies on nonlinear optical metasurfaces have primarily dealt with plasmonic effects produced in arrays of metallic nano-elements that support surface plasmon resonances, also called plasmonic nonlinear metasurfaces \cite{Krasnok2017,Panoiu2018,Rahimi2018}. Despite the significant local field enhancement effect that is observed in such plasmonic nanostructures, their use in nanophotonic applications has been limited by inherent thermal heating and large optical losses. This results in low damage thresholds at optical frequencies, which one seeks to avoid in practical photonic applications. More recently, the focus of research pertaining to nonlinear optical metasurfaces has gradually shifted to all-dielectric metasurfaces made of high-refractive-index materials \cite{Sain2019,Vabi2023,gigli2022}. These optical nanostructures present key advantages, including low dissipative losses and large damage threshold, thus allowing for larger operating optical pump power as compared to their plasmonic counterparts.

The basic components of all-dielectric metasurfaces are optical nanoresonators of various shape and different material properties that support Mie resonances \cite{Shche2014,Grin2017,Kivshar2022,Babi2024} of both electric and magnetic nature. This enables optical modes whose optical field is enhanced and strongly confined, although to a lesser extent when compared with plasmonic resonances of metallic nanoparticles. This is particularly relevant to nonlinear optical applications whereby dielectric materials with large optical nonlinearities are employed. In addition, the periodic nature of these metasurfaces facilitates coherent optical coupling among individual nanoresonators, giving rise to so-called guided-mode resonances (GMRs) \cite{Vabi2023,Quara2018,Huang2023GMR}. As a result, further enhancement of nonlinear optical effects can be achieved upon the excitation of such GMRs.

In addition to GMRs, optical bound-states in the continuum (BICs) provide a promising alternative for enhancing and controlling nonlinear optical processes in dielectric metasurfaces \cite{Hsu2016,Koshelev2019rev,Kang2023,Xu2023,Zeng2024}. More specifically, BICs are optical modes located inside the continuous part of the spectrum but spatially localized and completely decoupled from the radiation continuum. Consequently, they do not emit radiation into the far-field. Due to these and other unique optical properties, BICs have found a plethora of practical nanophotonics applications, ranging from sensing \cite{Tittl2018,Yesil2019,Chen2020} and lasing \cite{kodi2017,Huang2020laser,Han2023,Zhong2023,Chai2024} to nonlinear optics \cite{Koshelev2019THG,Minkov2019,Koshelev2020,Zograf2022,Cotru2023,Wang2024,Xu2019}. In particular, advances in harnessing frequency conversion processes driven by nonlinear BIC metasurfaces have brought about novel perspectives to a variety of device applications to modern technologies, including nonlinear imaging nanodevices \cite{Zheng2023,Xu2019}, quantum light sources \cite{San2022,Parry2021,Zhang2022SPDC}, non-reciprocal optical devices \cite{Cotru2023}, terahertz emitters \cite{Hu2022}, and chiral-optical nanodevices \cite{Shi2022,Koshelev2023}.

In general, an open extended periodic optical system supports two types of BICs: at-$\Gamma$ BICs (or symmetry-protected BICs), which arise from symmetry mismatch between a specific eigenmode (BIC) of the optical system and the available radiative channels, and off-$\Gamma$ BICs (or accidental BICs), which are usually achieved through precise tuning of the structural parameters of the optical system. Both mechanisms can lead to complete suppression of emission into all allowed radiation channels of the continuum. Because of this property, BICs can be viewed as optical resonances with infinite quality-($Q$) factor and lifetime.

In practical devices, BICs are realized in the form of quasi-BICs with finite $Q$-factor due to material absorption, finite periodic structure, surface roughness, and other perturbations \cite{Koshelev2019rev,Sadrieva2017}. For example, so far the largest reported value of an experimentally measured $Q$-factor of a BIC was about $10^4$ \cite{Liu2019,Huang2023Q,Zhong2024}. It should also be mentioned that recently a connection between BICs and Fano resonances has been established \cite{Koshelev2018}, which also explains the asymmetric line shape of BICs in transmission/reflection spectra.

Importantly, the topological nature of optical BICs was proposed and demonstrated by choosing a PhC slab as a case study \cite{Zhen2014}. Thus, it was proved that both types of BICs are vortex centers of the polarization of the far-field. As a consequence, the underlying physics of generation, evolution, and annihilation of BICs can be understood by introducing the concept of topological charge of the optical field, defined through the winding properties of the polarization vector in the \textbf{k}-space. Along with the enhanced light-matter interaction facilitated by BICs, the transition from vortex center to circularly polarized state by breaking the system symmetry \cite{Liu2019circular,Yoda2020,Gork2020,Chen2023nature} and the generation of vortex beams using the winding properties of the polarization vector centered at BICs \cite{Doeleman2018,Wang2020vortex,Wang2020optica,Kang2021AOM} provide new routes towards the investigation of nonlinear optical phenomena with intriguing properties.

Although both GMRs and BICs supported by dielectric metasurfaces enable enhanced light-matter interaction and strong local field confinement, the utilization of BICs-inspired metasurfaces for nonlinear optical applications provides several notable advantages: \textit{1)} optical BICs are characterized by practically infinite lifetime (a vanishingly amount of radiation is emitted into the far-field), whereas GMRs, even in the ideal case of no losses, can only possess a finite lifetime; \textit{2)} BICs exist in the momentum-space as vortex centers of far-field polarization, giving rise to the possibility to produce chiral effects and generate optical vortices, a functionality that GMRs do not provide; \textit{3)} symmetry-protection properties of BICs allow additional control of the line-width (or $Q$-factor) of induced resonances and enable optical states with an intricate structure of their polarization.
\begin{figure}[t]
\centering
\includegraphics[width=\columnwidth]{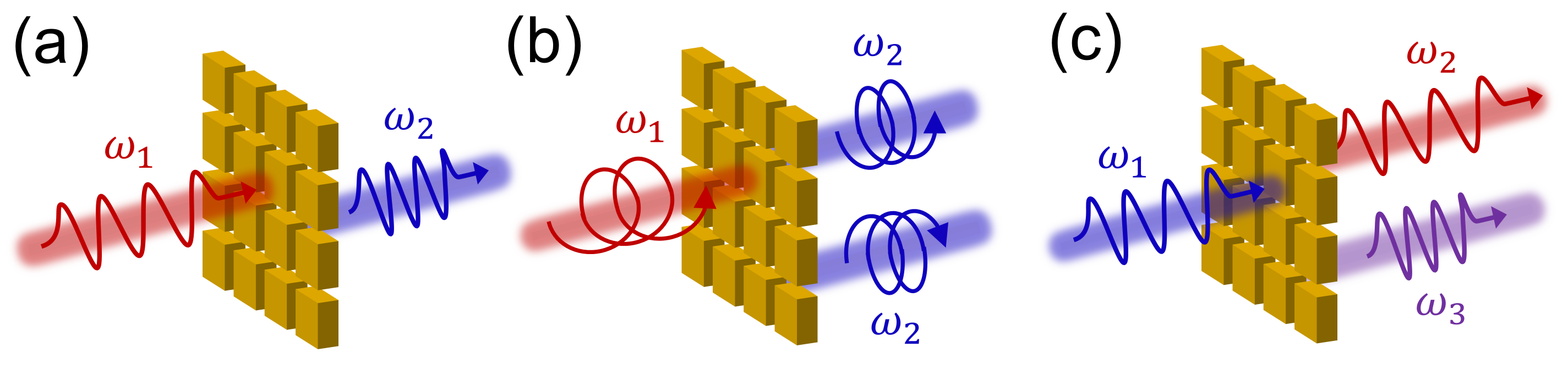}
\caption{Illustration of nonlinear optical metasurfaces based on optical BICs with different functionality. (a) Optical harmonic generation. (b) Nonlinear harmonic chiroptical effects. (c) Complex nonlinear optical frequency mixing; specifically, degenerate FWM is illustrated here, whereby $2\omega_{1}=\omega_{2}+\omega_{3}$.}\label{fig1}
\end{figure}

In what follows, we survey most relevant recent advances in the rapidly developing field of nonlinear optical metasurfaces that facilitate intense light-matter interaction \textit{via} the excitation of optical BICs. It should be mentioned that there are several excellent recent reviews of the field of optical BICs \cite{Koshelev2019rev,Kang2023,Xu2023,Zeng2024,Huang2023review,Wang2024review}, but they are primarily focused on linear optical properties and applications of BICs. We start our review with a brief introduction to the theoretical background pertaining to nonlinear optics, with special attention being paid to the second-order nonlinear optical effects and the physical concept of BICs supported by 2D periodic photonic structures. In particular, the topological nature of optical BICs for both symmetry-protected and accidental BICs and the multipole analysis of the the distribution of their optical field are also briefly discussed. We follow this introductory part with a discussion of the recent research developments related to metasurface-based nonlinear optical functionalities and applications in which BICs play a major role. These include harmonic generation, as per Fig.~\ref{fig1}(a),harmonic generation from dual BICs, nonlinear resonant chiral optical effects, illustrated in Fig.~\ref{fig1}(b), and complex nonlinear optical frequency mixing shown in Fig.~\ref{fig1}(c). Finally, the main conclusions of our review and a brief outlook on several potentially important future developments in BIC-related research areas are summarized in the last section.

\section{Theoretical Background}\label{sec2}
In this section, we introduce some of the fundamental physical concepts pertaining to nonlinear optical metasurfaces empowered by bound-states in the continuum. Specifically, we first describe some of the main nonlinear optical processes encountered in nonlinear optics and afterwards we introduce and discuss the optical properties of BICs, with a particular focus being placed on their nontrivial topological nature.

\subsection{Nonlinear optical processes}
Nonlinear optics is the branch of science that describes the interaction between light and matter in nonlinear optical media, in which the optical response of an optical material depends nonlinearly on the applied electric field. To describe such nonlinear optical interactions, the electric polarization, $\textbf{P}(\textbf{r},t)$, induced within a medium is expanded as a power series with respect to the electric field, $\textbf{E}(\textbf{r},t)$, under the electric-dipole approximation \cite{Boyd2008,Panoiu2024},
\begin{align}
\begin{split}
    \textbf{P}(\textbf{r},t)&=\epsilon_0\chi^{(1)} \cdot \textbf{E}(\textbf{r},t)+\epsilon_0\chi^{(2)}: \textbf{E}(\textbf{r},t)\textbf{E}(\textbf{r},t)+ \epsilon_0\chi^{(3)}\vdots \textbf{E}(\textbf{r},t)\textbf{E}(\textbf{r},t)\textbf{E}(\textbf{r},t)+\dots \\
    &\equiv\textbf{P}^{(1)}(\textbf{r},t)+ \textbf{P}^{(2)}(\textbf{r},t)+ \textbf{P}^{(3)}(\textbf{r},t)+\dots,
    \label{Eq1}
\end{split}
\end{align}
where $\chi^{(1)}$ is the linear electric susceptibility and $\chi^{(2)}$ and $\chi^{(3)}$ are the second- and third-order nonlinear optical susceptibilities, respectively. Typically, $\chi^{(1)}$ is a second-rank tensor, $\chi^{(2)}$ is a third-rank tensor, and so on. The tensor property of susceptibilities depends on the crystal structure of a material, whereas for homogeneous and isotropic materials the susceptibilities are scalar quantities. Regarding the polarization defined by Eq.~(\ref{Eq1}), one can divide it into linear and nonlinear parts: the linear part $\textbf{P}^{(1)}(\textbf{r},t)$ describes the conventional or linear optics, whereas $\textbf{P}^{(2)}(\textbf{r},t)$, $\textbf{P}^{(3)}(\textbf{r},t)$, and any other higher-order polarizations describe different nonlinear optical processes.

As one of the most important nonlinear optical phenomena, we consider here in more detail the second-order nonlinear process in a nonlinear medium characterized by $\chi^{(2)}$. To this end, we assume that the incident field is a superposition of two monochromatic plane waves:
\begin{equation}
    \textbf{E}(\textbf{r},t)=\textbf{E}_1 e^{i(\textbf{k}_1 \cdot \textbf{r}-\omega_1 t)}+\textbf{E}_2 e^{i(\textbf{k}_2 \cdot \textbf{r}-\omega_2 t)}+ c.c.,
\label{Eq2}
\end{equation}
where $\omega_1$ and $\omega_2$ represent the two different frequency components, $\textbf{E}_1$ and $\textbf{E}_2$ are the wave amplitudes, $\textbf{k}_1$ and $\textbf{k}_2$ are the wave vectors, and $c.c.$ signifies complex conjugation. Next, using Eq.~(\ref{Eq2}), the second-order nonlinear polarization in Eq.~(\ref{Eq1}) can be cast in the form of
\begin{align}
\begin{split}
    \textbf{P}^{(2)}(\textbf{r},t)=& \epsilon_0 \chi^{(2)}:[\textbf{E}_1\textbf{E}_1 e^{2i(\textbf{k}_1 \cdot \textbf{r}-\omega_1 t)}+ \textbf{E}_2\textbf{E}_2 e^{2i(\textbf{k}_2 \cdot \textbf{r}-\omega_2 t)} \\
    & + 2\textbf{E}_1\textbf{E}_2 e^{i[(\textbf{k}_1+\textbf{k}_2) \cdot \textbf{r}-(\omega_1+\omega_2) t]} +2\textbf{E}_1\textbf{E}_2^* e^{i[(\textbf{k}_1-\textbf{k}_2) \cdot \textbf{r}-(\omega_1-\omega_2) t]}+ c.c.] \\
    & + 2\epsilon_0 \chi^{(2)}:(\textbf{E}_1\textbf{E}_1^*+ \textbf{E}_2\textbf{E}_2^*).
\label{Eq3}
\end{split}
\end{align}

For convenience, we decompose this nonlinear polarization as follows:
\begin{equation}
    \textbf{P}^{(2)}(\textbf{r},t)= \sum_n \textbf{P}^{(2)} (\omega_n) e^{-i\omega_n t}.
\label{Eq4}
\end{equation}
Then, the frequency components in the second-order nonlinear polarization are given by
\begin{subequations}
\label{Eq5}
\begin{align}
    & \textbf{P}^{(2)}_{\mathrm{SHG}}(2\omega_1)=\epsilon_0 \chi^{(2)}:\textbf{E}_1\textbf{E}_1,
    \label{Eq5a} \\
    & \textbf{P}^{(2)}_{\mathrm{SHG}}(2\omega_2)=\epsilon_0 \chi^{(2)}:\textbf{E}_2\textbf{E}_2,
    \label{Eq5b} \\
    & \textbf{P}^{(2)}_{\mathrm{SFG}}(\omega_1+\omega_2)=2\epsilon_0 \chi^{(2)}:\textbf{E}_1\textbf{E}_2,
    \label{Eq5c} \\
    & \textbf{P}^{(2)}_{\mathrm{DFG}}(\omega_1-\omega_2)=2\epsilon_0 \chi^{(2)}:\textbf{E}_1\textbf{E}_2^*,
    \label{Eq5d} \\
    & \textbf{P}^{(2)}_{\mathrm{OR}}(0)=2\epsilon_0 \chi^{(2)}:(\textbf{E}_1\textbf{E}_1^*+\textbf{E}_2\textbf{E}_2^*).
    \label{Eq5e}
\end{align}
\end{subequations}
Here, SFG, DFG, and OR denote sum-frequency generation, difference-frequency generation, and optical rectification, respectively. It should be noted that the terms with (unphysical) negative frequency are omitted from Eq.~(\ref{Eq4}). Commonly used noncentrosymmetric dielectric materials for even-order nonlinear processes in all-dielectric metasurfaces include gallium arsenide (\ch{GaAs}) \cite{Liu2016,Vabi2018}, aluminum gallium arsenide (\ch{AlGaAs}) \cite{Cama2016,Marino2019}, lithium niobate (\ch{LiNbO3}) \cite{Wang2017,Zhang2022,Boes2023}, monolayer transition metal dichalcogenides (TMDs) \cite{Bernhar2020,Lochner2020,Huang20242D,Wang20242D,Ren2024}, and 2D gallium selenide (GaSe) \cite{Huang20242D,Liu20212D}.

Likewise, higher-order nonlinear processes can also be described in a similar way as the second-order one is described by Eq.~(\ref{Eq3}). For the general case in which the input consists of several different frequency components, the expression for $\textbf{P}^{(3)}(\textbf{r},t)$ is quite lengthy so that it is not given here. Instead, we present the most common cubic nonlinear optical interaction in nonlinear metasurfaces governed by BICs, namely THG, which describes the generation of a photon at frequency $3\omega$ from three photons at frequency $\omega$:
\begin{equation}
    \textbf{P}^{(3)}_{\mathrm{THG}}(3\omega)=\epsilon_0 \chi^{(3)} \vdots \textbf{E}(\omega
    )\textbf{E}(\omega)\textbf{E}(\omega).
\label{Eq6}
\end{equation}

Regarding the dielectric materials used for odd-order nonlinear optical processes, amorphous silicon (a-Si) \cite{Koshelev2019THG,Wang2024,Xu2019,Tang2024,Xu2020}, crystalline silicon (c-Si) \cite{Liu2019,Gao2018}, GaAs \cite{Liu2018FWM,Genna2022}, and amorphous germanium (a-Ge) \cite{Zubyuk2022,Yeze2023} are most commonly employed.

Before we conclude this section, we discuss SHG optical process in centrosymmetric optical media. Under the electric dipole approximation, SHG is forbidden in a medium whose atomic structure possesses inversion symmetry. Here, we introduce a widely used model to describe the second-order nonlinear optical process from centrosymmetric optical media that goes beyond the electric dipole approximation. In particular, the model accounts for SHG from both higher-order (nonlocal) effects inside the bulk of the nonlinear medium and dipole-allowed (local) surface response at the interface between the centrosymmetric material and the environment where the inversion symmetry is broken \cite{Heinz1991}. The surface and bulk contributions to the nonlinear optical response at the second-harmonic from centrosymmetric media can thus be described \textit{via} two nonlinear polarizations \cite{Panoiu2018,Heinz1991}:
\begin{subequations}
\label{Eq7}
\begin{align}
    \textbf{P}^{(2)}_{s}(2\omega;\textbf{r}) = & \epsilon_0 \textbf{$\chi$}^{(2)}_{s} : \textbf{E}(\omega;\textbf{r}) \textbf{E}(\omega;\textbf{r}) \delta(\textbf{r}-\textbf{r}_s),
    \label{Eq7a} \\
\begin{split}
    \textbf{P}^{(2)}_{b,i}(2\omega;\textbf{r}) = & \gamma\nabla_i[\textbf{E}(\omega;\textbf{r}) \cdot \textbf{E}(\omega;\textbf{r})] + \beta[\nabla \cdot \textbf{E}(\omega;\textbf{r})]E_i(\omega;\textbf{r}) \\
    & + \zeta E_i(\omega;\textbf{r})\nabla_i E_i(\omega;\textbf{r}) + \delta^{\prime} [\textbf{E}(\omega;\textbf{r}) \cdot \nabla] E_i(\omega;\textbf{r}),
    \label{Eq7b}
\end{split}
\end{align}
\end{subequations}
where $\textbf{P}^{(2)}_{s}$ and $\textbf{P}^{(2)}_{b}$ denote the surface and bulk quadratic nonlinear polarizations, respectively, Dirac delta-function $\delta(\textbf{r}-\textbf{r}_s)$ defines the surface of the nonlinear material, $\chi^{(2)}_{s}$ is the surface nonlinear susceptibility tensor, and $\gamma$, $\beta$, $\zeta$, and $\delta^{'}$ are material parameters that characterize the (nonlocal) electric quadrupole and magnetic dipole contributions to the bulk part of nonlinear polarization.

Due to the isotropic mirror symmetric property at the interface of the nonlinear medium, the second-order surface susceptibility tensor $\chi^{(2)}_{s}$ has only three non-vanishing independent components: $\chi^{(2)}_{s,\perp\perp\perp}$, $\chi^{(2)}_{s,\perp\parallel\parallel}$, and $\chi^{(2)}_{s,\parallel\perp\parallel} = \chi^{(2)}_{s,\parallel\parallel\perp}$, where the symbol $\perp$ ($\parallel$) denotes the normal (tangential) direction with respect to the interface. Regarding the bulk part of nonlinear polarization in Eq.~(\ref{Eq7b}), the second term vanishes as $\nabla \cdot \textbf{E}(\omega;\textbf{r})=0$ in homogeneous media. In addition, the $\zeta$ parameter in the third term depends on the anisotropy of materials. Concerning the $\delta^{\prime}$ term in Eq.~(\ref{Eq7b}), theoretical models predict that it can be neglected, too \cite{Sipe1980}.

\subsection{Bound states in the continuum}
\begin{figure}[!t]
\centering
\includegraphics[width=\columnwidth]{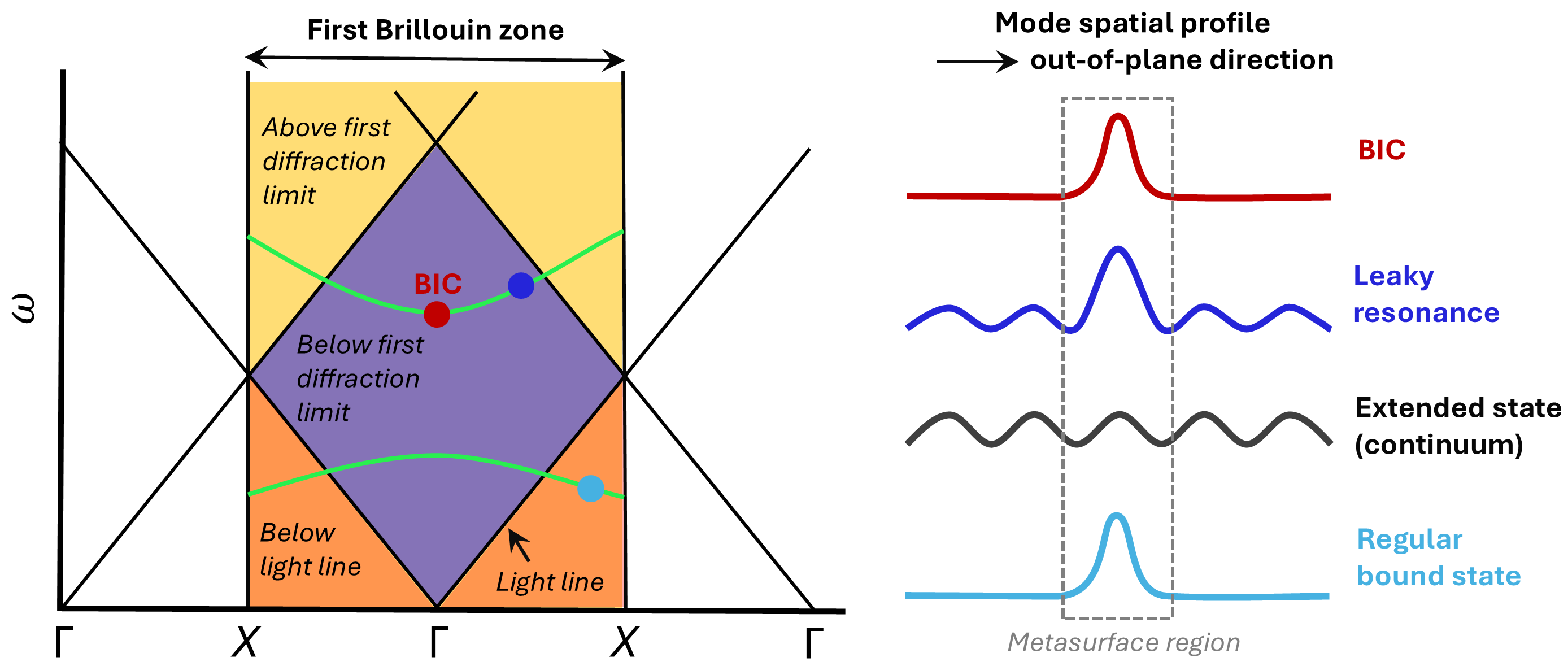}
\caption{Illustration of a BIC of a 2D square lattice metasurface lying in the $x-y$-plane. Left panel: schematic of the diffraction cutoff frequencies and partition of the first Brillouin zone at $k_y = 0$ for homogeneous and isotropic surrounding medium, with two typical photonic bands of the metasurface depicted by green curves. The regions in orange are below the light line, the region in purple is below the first Bragg-diffraction limit and above the light line, and the regions in yellow are above the first Bragg-diffraction limit. $\Gamma$ and $X$ symmetry points correspond to $(k_x,k_y) = (0,0)$ and $(k_x,k_y) = (\pi/a,0)$, respectively, where $a$ is the lattice constant. Right panel: spatial profiles along the out-of-plane direction of the metasurfaces of several generic modes, namely a regular bound state (light blue circle in the left panel), an extended state (modes spanning the regions in purple and yellow in the left panel), a leaky resonance (dark blue circle in the left panel), and a BIC (red circle in the left panel).}\label{fig2}
\end{figure}

Optical BICs refer to optical states possessing the unusual property of being completely localized despite the fact that their frequency lies within the continuum part of the spectrum of an open optical system. In other words, BICs are discrete, localized states embedded in a continuous spectrum consisting of extended, propagating waves, but completely decoupled from them. To illustrate and explain the physical properties of BICs, let us consider the band diagramand mode spatial profile along the out-of-plane direction of a 2D metasurface consisting of a periodic square array of basic elements, as per Fig.~\ref{fig2}.

Within the spectral region below the light line (orange region), there exist regular bound states (light blue circle) whose frequencies form a discrete spectrum. As they undergo total internal reflection, these optical states are confined within the plane of the metasurface (see the right panel of Fig.~\ref{fig2}) and can only propagate within the plane of the metasurface. The light line is given by the dispersion curve $\omega = ck_{\parallel}/n$, where $\omega$ is the frequency, $c$ is the speed of light in vacuum, $k_{\parallel}$ is the amplitude of the component of the Bloch wavevector lying in the plane of the metasurface, and $n$ is the refractive index of the surrounding environment.

For modes above the light line and below the first Bragg-diffraction limit (purple region in Fig.~\ref{fig2}), only a pair of radiative modes with orthogonal polarization ($s$- or $p$-polarized) can exist. Therefore, generally, these radiative modes can couple with the states in the continuum (regions colored in purple and yellow, defined by $\omega > ck_{\parallel}/n$) under the condition of wavevector matching and nonvanishing field overlap, giving rise to a resonant mode profile in the vicinity of metasurfaces with energy leakage into the surrounding environment (see the right panel of Fig.~\ref{fig2}). As a result, so-called leaky resonances marked by a dark blue circle in Fig.~\ref{fig2} and characterized by a complex frequency, $\omega=\omega_0-i\gamma$, can exist, where $\omega_0$ represents the resonance frequency and $\gamma$ is the damping rate of the resonance; they are related to the $Q$-factor \textit{via} $Q=\omega_0/(2\gamma)$. However, it is possible that within this spectral range of the photonic band diagram there exist certain specific states, called BICs (marked by a red circle), defined by $\gamma=0$, which are completely decoupled from the surrounding continuum \textit{via} symmetry protection or precise tuning of the geometrical parameters of the metasurface. Thus, BICs are characterized by an out-of-plane spatially confined optical field, as illustrated in the right panel of Fig.~\ref{fig2}. Despite being located within the continuum part of the spectrum, these symmetry-protected and accidental BICs possess peculiar properties, markedly different from those of leaky modes.

Above the first-order Bragg-diffraction limit (yellow region), more radiative channels consisting of different diffraction beams open, which makes it impossible to simultaneously cancel the optical coupling between the radiation continuum and all the diffraction beams contained in a symmetry-protected BIC \cite{Cerjan2021}. As a result, symmetry-protected BICs cannot exist above the first-order diffraction limit, so that most of the studies pertaining to optical BICs so far have been focused on the spectral region below the first-order diffraction limit.

It was demonstrated that the topological properties of BICs were intimately related to the flow in the momentum  ($\textbf{k}$) space of the polarization vector of the far-field \cite{Zhen2014}, namely they can be viewed as vortex centers in the $\textbf{k}$-space. Specifically, if one considers a 2D metasurface that is periodic along $x$- and $y$-direction, one can express the electric field of the BIC using the Bloch theorem as follows: $\textbf{E}_{\textbf{k}_{\parallel}} (\bm{\rho},z)=e^{i\textbf{k}_{\parallel}\cdot\bm{\rho}} \textbf{u}_{\textbf{k}_{\parallel}}(\bm{\rho},z)$, where $\bm{\rho}=x\hat{x}+y\hat{y}$ is the in-plane position vector, $z$ is the coordinate along the direction normal onto the metasurface, $\textbf{k}_{\parallel}=k_x\hat{x}+k_y\hat{y}$ is the Bloch wave vector, and $\textbf{u}_{\textbf{k}_{\parallel}}$ is a periodic vector function.

Below the first-order diffraction limit and above the light line, the only nonzero amplitude of propagating plane waves is the zero-order Fourier component of $\textbf{u}_{\textbf{k}_{\parallel}}$, given by $\textbf{c}_{\textbf{k}_{\parallel}}=c_x(\textbf{k}_{\parallel})\hat{x}+c_y(\textbf{k}_{\parallel})\hat{y}$. Here, $c_x(\textbf{k}_{\parallel})$ and $c_y(\textbf{k}_{\parallel})$ are defined as the $x$- and $y$-component of the spatial average of $\textbf{u}_{\textbf{k}_{\parallel}}$ over a unit cell across any horizontal plane outside the metasurface, denoted as $\braket{\textbf{u}_{\textbf{k}_{\parallel}}}$. Then, the vortex behavior of the far-field polarization in the momentum space can be quantitatively characterized by introducing the topological charge, $q$, carried by a BIC:
\begin{equation}
    q=\frac{1}{2\pi}\oint_{\mathcal{C}} d\textbf{k}_{\parallel} \cdot \nabla_{\textbf{k}_{\parallel}}\phi(\textbf{k}_{\parallel}).
    \label{Eq8}
\end{equation}
Here, $\mathcal{C}$ represents a closed $\textbf{k}$-path that encircles the BIC along the counterclockwise direction and $\phi(\textbf{k}_{\vert \vert})= \mathrm{arg}[c_x(\textbf{k}_{\vert \vert})+ic_y(\textbf{k}_{\vert \vert})]$ is the polarization angle. The topological charge determines the number of times that the polarization vector winds around the BIC along the closed \textbf{k}-path $\mathcal{C}$.

A powerful technique particularly useful to analyze the formation mechanisms of both symmetry protected and accidental non-radiating BICs in 2D periodic structures is provided by the multipole expansion technique. Within the framework of this method, the radiation field in the far-field region of an isolated dielectric meta-atom is decomposed into different orders of electromagnetic multipoles and used in conjunction with the lattice symmetry properties to find out the directions along which radiation cannot be emitted \cite{Sadri2019,Glady2022}.

To conclude this section we note that we have provided only a brief introduction to some of the fundamental concepts pertaining to the topological nature of BICs. An interested reader can learn more about additional important concepts, such as the lattice-symmetry-dependent topological charges, merging BICs, and split of BICs into circularly polarized states, from several excellent sources, such as Refs.~\cite{Kang2023,Zhen2014,Yoda2020,Kang2021,Kang2022}.

\section{BIC-assisted Nonlinear Light-matter Interactions in Metasurfaces}\label{3}
The effect of enhancement of the local optical field \textit{via} the excitation of BICs can be harnessed to improve and manipulate the frequency conversion efficiency of nonlinear optical interactions mediated by optical metasurfaces. Moreover, due to the topological nature of BICs, nonlinear optical metasurfaces can also be employed to achieve highly effective nonlinear chiroptical effects characterized by large conversion efficiency. Starting from these considerations, in this section we review some recent advances in harmonic generation, nonlinear chiroptical effects, and complex frequency mixing processes empowered by BIC-resonant nonlinear optical metasurfaces.

\subsection{Harmonic generation}\label{3.1}
Among the multitude of applications of BICs relying on the nonlinear optical response of planar optical metasurfaces, harmonic generation is perhaps the most thoroughly investigated \cite{Koshelev2019rev,Kang2023,Xu2023,Zeng2024}. It represents the generation of photons with frequency equal to an integer multiple of that of the incoming photons, whereby the nonlinear optical material mediates the optical interaction. For instance, SHG processes produce photons with frequency twice as large as the frequency of the incoming photons, the photon frequency is tripled upon THG optical interactions, and so on.

Generally, the strength of nonlinear optical interactions weakens when the order of the nonlinear process increases \cite{Boyd2008}. Consequently, one needs to apply higher-intensity pump optical powers for efficient generation of nonlinear signals at harmonics of increasing order. Their key feature of forming high-$Q$ optical resonances and realizing extremely confined and enhanced local optical near-field, in conjunction with the high damage power thresholds of dielectric materials, render all-dielectric metasurfaces ideal platforms for implementing nonlinear optics applications. In particular, enhanced nonlinear harmonic generation from BIC-assisted metasurfaces has been intensively studied over the last few years, with demonstration of nonlinear optical interactions with order ranging from second-order up to eleventh-order being reported \cite{Vabi2018,Zheng2023,Zogra2022,Koshelev2019THG,Wang2024,Xu2019,Liu2019,Zhang2022,Bernhar2020,Liu20212D,Xiao2022,Anthur2020,Fang2022,Abde2024,Fan2024}.

\begin{figure}[!t]
\centering
\includegraphics[width=\columnwidth]{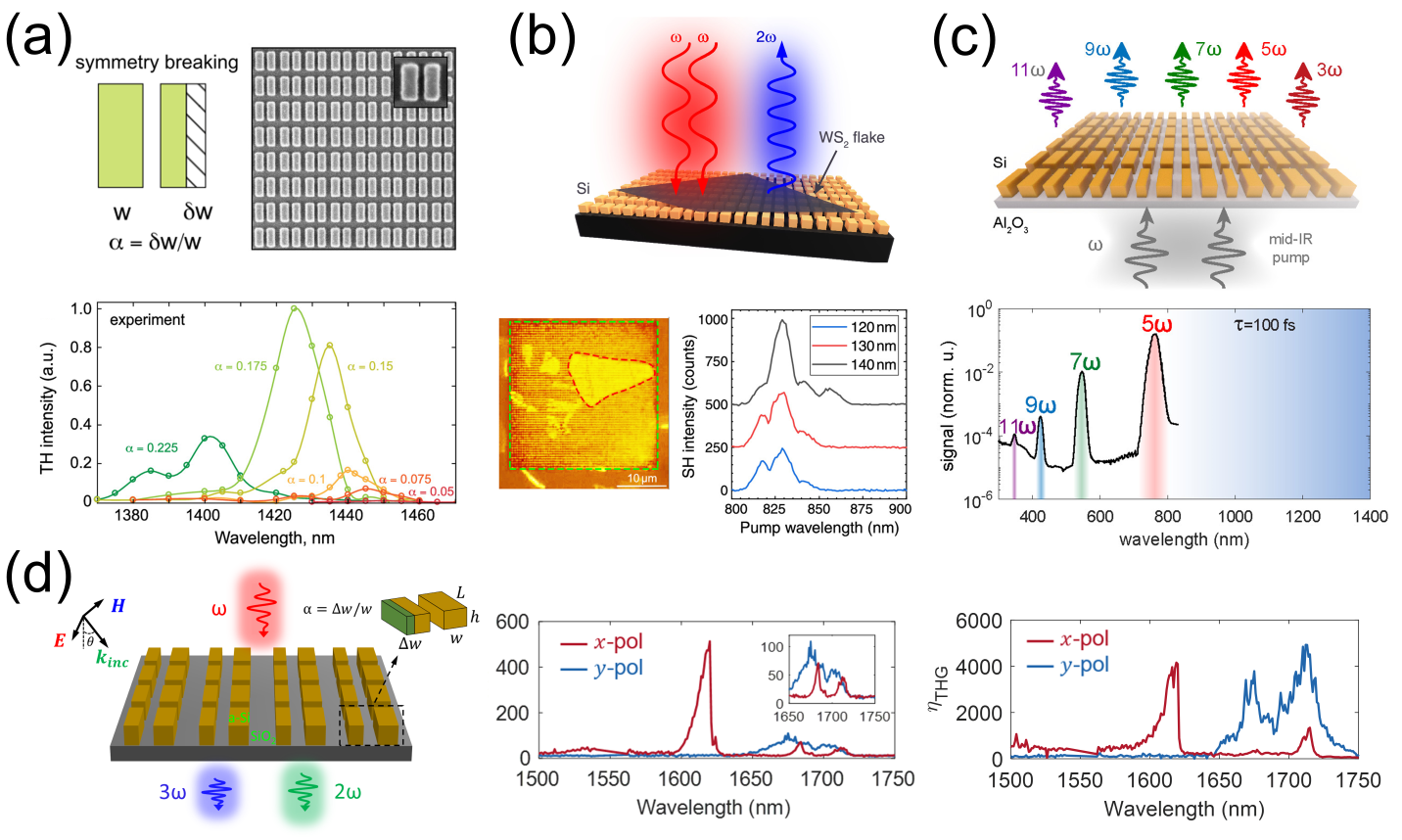}
\caption{Examples of enhanced nonlinear harmonic generation in all-dielectric metasurfaces associated with BIC-type resonances. (a) Symmetry-broken Si metasurface for tailoring and boosting THG. Top left panel, a meta-atom with broken in-plane inversion symmetry characterized by asymmetry parameter, $\alpha$. Top right panel, SEM of the resonant metasurface. Bottom panel, evolution of measured THG with respect to geometry asymmetry. Reproduced with permission from \cite{Koshelev2019THG}. (b) Resonant enhancement of SHG from monolayer \ch{WS2}. Top panel, schematic of the hybrid photonic structure composed of a \ch{WS2} monolayer placed onto a Si metasurface. Bottom left panel, optical microscopy image with green and red dashed lines showing Si structure and monolayer \ch{WS2}, respectively. Bottom right panel, measured SH intensity for three different metasurface samples. Reproduced with permission from \cite{Bernhar2020}. (c) Odd-order high-harmonic generation empowered by BIC-resonant metasurfaces. Top panel, concept of asymmetric Si metasurface. Bottom panel, spectrum of harmonics generated at $5\omega$, $7\omega$, $9\omega$, and $11\omega$ by \SI{100}{\fs} laser pulses. Reproduced with permission from \cite{Zograf2022}. (d) Enhanced SHG and THG from symmetry-broken metasurfaces made of (centrosymmetric) a-Si. Left panel, schematic representation of the nonlinear metasurface. Middle panel, measured enhancement effect of SHG from both local surface and nonlocal bulk contributions. The inset shows the zoom-in data. Right panel, experimentally observed enhancement factor of THG. Reproduced with permission from \cite{Wang2024}.}\label{fig3}
\end{figure}

One general approach to generate strong nonlinear interactions in metasurfaces was recently proposed \cite{Koshelev2019THG}, whereby the sharp resonances associated with BICs have been employed. It was experimentally and numerically demonstrated that THG from metasurfaces comprising symmetry-broken Si nanoresonators, as per Fig.~\ref{fig3}(a), can be engineered and enhanced by tuning the asymmetry parameter, $\alpha = \delta w/w$. This allows one to vary the radiative $Q$-factor, $Q_r$, of the quasi-BIC, according to the relation $Q_r = Q_0/ \alpha^{2}$ \cite{Koshelev2018}, where $Q_0$ is a metasurface design-dependent constant. One immediate consequence of this fact is that, using the time-domain coupled-mode theory \cite{Chris2024}, one can show that the maximum intensity of the TH signal is achieved in the critical coupling regime, namely when the radiative loss into the available diffraction channel, $Q_r$, and the nonradiative loss, $Q_{nr}$, arising from material absorption, surface scattering, and fabrication imperfections, exactly balance each other, that is $Q_r = Q_{nr}$. This feature is illustrated in the bottom panel of Fig.~\ref{fig3}(a). In particular, the maximum conversion efficiency, $P_{3\omega}/P_{\omega}$, was estimated to be about $10^{-6}$ for the average pump power of $P_{\omega}=\SI{130}{\milli\watt}$.

By operating a BIC-based nonlinear metasurface in the critical coupling regime one can also boost the effective nonlinear susceptibility of 2D layered materials, such as TMDs monolayer \cite{Bernhar2020,Lochner2020,Wang20242D,Ren2024}. These 2D materiale possesses second-order nonlinearity when they contain an odd-number of layers, with the maximum quadratic nonlinearity being achieved in a single-layer configuration \cite{Li2013,Zeng2013,Weismann2016,You2018}. Although 2D materials exhibit large surface nonlinear susceptibility, the short interaction distance with light greatly limits the intensity of the nonlinear optical response. To overcome this difficulty, one can employ optical resonances of the photonic nanostructure onto which the 2D material is placed.

In this context, as per Fig.~\ref{fig3}(b), it has been demonstrated that by exploiting the interplay between the radiative and nonradiative losses, the intensity of the SHG in a tungsten disulfide (\ch{WS2}) monolayer placed on top of a Si metasurface possessing BICs can be increased by at least 1140 times as compared to the case when the 2D material is placed onto a flat, unstructured substrate \cite{Bernhar2020}. In addition, it has been shown that 2D gallium selenide (GaSe) flakes can couple with quasi-BICs supported by a Si metasurface under continuous-wave pump conditions, resulting in a giant enhancement of SHG of up to 9400 times \cite{Liu20212D}. These examples of successful integration of 2D materials with BIC-inspired dielectric metasurfaces point to novel applications for optoelectronics and nonlinear optical microscopy.

The large $Q$-factor and field confinement effect associated to the excitation of symmetry-protected BICs can also facilitate the generation of high harmonics at the nanoscale. This idea is illustrated by the schematic of an asymmetric Si metasurface presented in Fig.~\ref{fig3}(c), an optical metasurface that supports a BIC-type resonance in the mid-IR part of the spectrum \cite{Zograf2022}. The metasurface is designed so as to enhance nonlinear optical signals arising from odd-order nonlinear processes. Under illumination with \SI{100}{\fs} laser pulses, nonlinear harmonic signals ranging from $3^{\mathrm{rd}}$- to $11^{\mathrm{th}}$-order were experimentally observed. Importantly, going beyond the perturbative regime where the nonlinear light-matter interaction can be viewed as a small perturbation to its linear counterpart, the transition from perturbative to nonperturbative nonlinear optics was observed, thus opening up new avenues towards the exploration of the physics of nonperturbative optical nonlinearities at subwavelength scale \cite{Zograf2022,Sinev2021}.

In a different study, a polarization-controlled all-dielectric optical metasurface that possesses a pair of BICs (a symmetry-protected BIC and an accidental one) was proposed and numerically investigated. It has been demonstrated that it can be used to implement a dynamical switch of high-harmonic generation, including THG and fifth-harmonic generation (FHG). In this study, the FHG was calculated assuming both direct and cascaded processes \cite{Xiao2022}. This optical metasurface design provides a novel approach to achieve tunable and switchable nonlinear optical signals by simply taking advantage of the structural symmetry of the metasurface and the polarization of the optical pump.

For centrosymmetric optical media whose atomic structure is invariant upon inversion symmetry transformation, even-order nonlinear optical processes, such as SHG, are forbidden under the electric dipole approximation. Therefore, research in this area has primarily been focused on odd-order nonlinear optical processes in metasurfaces made of centrosymmetric optical media, such as silicon. However, due to the broken inversion symmetry at the interface between the nonlinear (centrosymmetric) optical medium and the surrounding environment, and the higher-order multipoles induced in the bulk of the medium, quadratic nonlinear response can be achieved in this case, too \cite{Wang2024,Liu2019,Fang2022}; see, also, Eq.~(\ref{Eq7}).

In a recent work \cite{Wang2024}, it was investigated both theoretically and experimentally the SHG and THG observed upon nonlinear interaction of light with an a-Si metasurface consisting of a periodic distribution of bar-shaped elements, as per Fig.~\ref{fig3}(d). One conclusion of this study was that the SHG intensity can be enhanced by almost three orders of magnitude, which is primarily due the field enhancement effects associated to the excitation of BICs and GMRs. In addition, this analysis revealed a 5000-fold maximum THG enhancement factor, attributable to the resonant optical response of the metasurface. According to the experimental measurements, the up-conversion efficiencies of second-order ($P_{2\omega}/P_{\omega}$) and third-order ($P_{3\omega}/P_{\omega}$) nonlinear processes were estimated to be $10^{-8}$ and $10^{-7}$, respectively. Furthermore, the diffractive properties of the nonlinear optical fields generated at the SH and TH were also investigated, namely the distribution of the optical power among the diffraction orders of the two nonlinear optical beams generated at the SH and TH.

\subsection{Harmonic generation from dual bound-states in the continuum}
\label{3.2}
\begin{figure}[!b]
\centering
\includegraphics[width=\columnwidth]{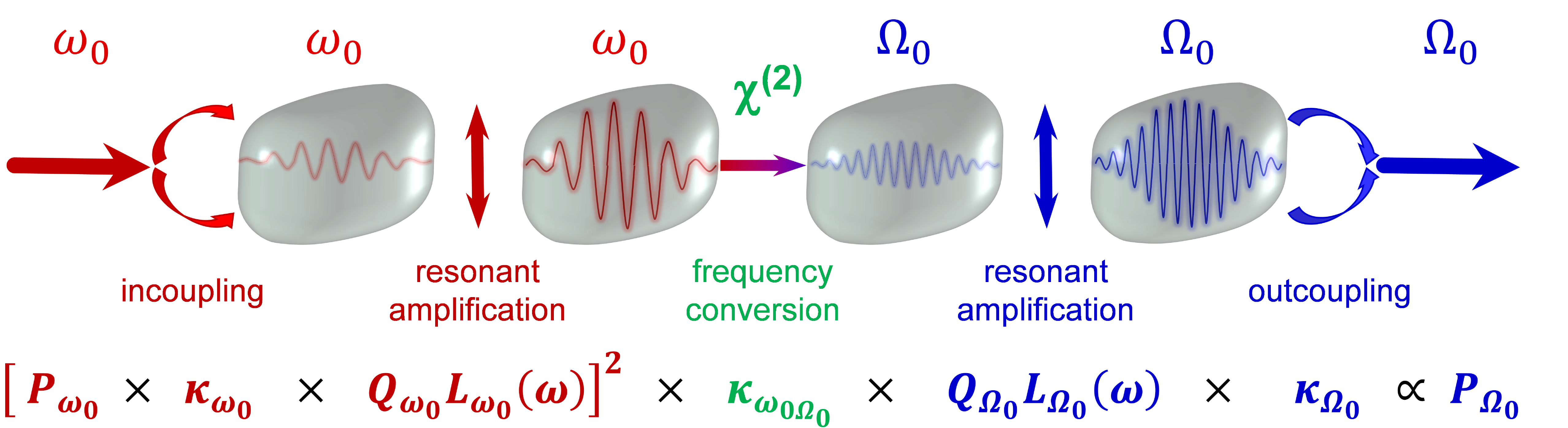}
\caption{Schematic of the SHG in an optical nanoresonator. Each term of the formula describes either a linear or nonlinear step of the nonlinear optical process. The incoming (FF) and outgoing (SH) light has frequency $\omega_{0}$ and $\Omega_{0}=2\omega_{0}$, respectively. Reproduced with permission from \cite{Panoiu2024}.}\label{fig4}
\end{figure}

The pursuit of efficient nonlinear frequency conversion lies at the center of the research in nonlinear optics of photonic nanostructures. To achieve this goal, a particularly effective approach is based on the so-called double-resonance phenomenon \cite{Lin2016,You2017,Qun2019,You2020,Lan2021}. Taking SHG as an example, this approach can be summarized as follows: First, the incoming light at the FF couples into the optical mode of the system and is resonantly amplified due to the high $Q$-factor of this mode. Then, optical power is transferred to the mode that exists at the SH, \textit{via} SHG interaction mediated by the quadratically nonlinear susceptibility of the optical resonator. The optical field in this mode at the SH is then resonantly enhanced due to the high $Q$-factor of the nonlinear mode. Finally, the optical power at the SH is outcoupled to radiative modes of the continuum. These ideas are schematically illustrated in Fig.~\ref{fig4}.

Equally important, the optical overlap between the two interacting resonances, mediated by the nonlinear optical susceptibility tensor of the medium, should be optimized in such a way that the efficiency of the frequency conversion process is maximized. Our discussion in the preceding section covered recent advances in the understanding of BIC-governed harmonic generation in dielectric optical metasurfaces. In all cases discussed, the nonlinear optical metasurface supports a BIC-type resonance only at the FF. The case that presumably has the potential to lead to a much larger conversion efficiency, namely when there are BICs at both interacting frequencies, has been much less investigated.
\begin{figure}[!b]
\centering
\includegraphics[width=\columnwidth]{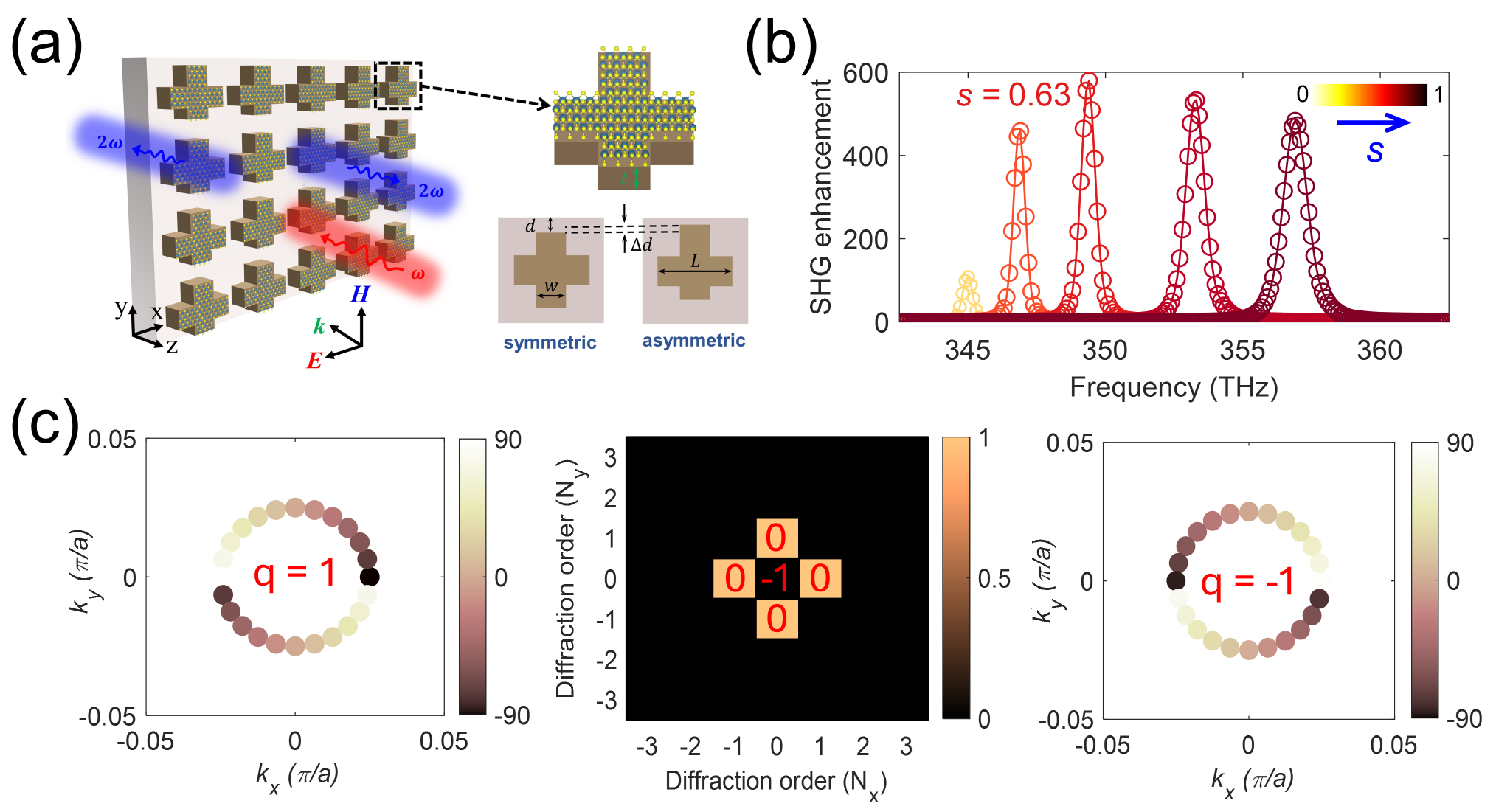}
\caption{Enhanced SHG in composite monolayer \ch{MoS2}/Si metasurfaces arising from excitation of a pair of quasi-BICs. (a) Conceptual diagram of a symmetry-broken (asymmetry parameter $s = \Delta d/d$) nonlinear metasurface with monolayer \ch{MoS2} placed on top of cruciform-shaped Si resonators. (b) Spectra of SHG enhancement factor determined for different metasurface asymmetry $s$. Maximum resonant enhancement effect is observed at $s=0.63$. (c) Topological analysis of interacting optical resonances in the case of symmetric meta-atoms. Left panel, evolution of far-field polarization angle of the zeroth-order diffraction channel along a closed \textbf{k}-path centered at the BIC at the FF, suggesting a topological charge $q=1$. Middle panel, Fourier analysis of the far-field of the BIC-type resonance at the SH with topological charge $q$ indicated for the five diffraction channels. Right panel, evolution of the zeroth-order diffraction channel of the SH resonance along a \textbf{k}-path, indicating a topological charge of -1. Reproduced with permission from \cite{Wang20242D}.}\label{fig5}
\end{figure}

Given practical considerations, the frequency of the resonance at the SH, which is one of the two components of the dual-resonance pair supported by a nonlinear optical metasurface, usually lies in the spectral region located above the first diffraction limit. A common practice, however, is to employ BICs that exist in the spectral region lying above the light line and below the first diffraction limit, represented by the purple regions in Fig.~\ref{fig2}. For such BICs, only the zero-order Fourier coefficient of the optical field needs to be engineered so as to achieve either symmetry-protected or accidental BICs \cite{Hsu2016,Kang2023,Xu2023,Zeng2024}.

For optical resonances with frequency above the first diffraction limit, there exist additional diffraction channels that can carry energy into the far-field and thus lower the radiative $Q$-factor. This makes it difficult to simultaneously and completely suppress the radiation into all existing diffraction channels. Accordingly, very little attention has been paid off to metasurfaces and photonic crystals that support BIC-like resonances with frequency above the first diffraction limit. However, even when such resonances are not completely decoupled from the radiation continuum, their $Q$-factor can be fairly large. This is achieved through similar techniques as those employed to construct true BICs, namely based on symmetry considerations and parameter tuning. In this way, it is possible to suppress radiation from some of the diffraction channels and thus design quasi-BICs with large $Q$-factor.

Recently, this novel approach was employed to enhance SHG in 2D TMD materials whereby the optical fields at the FF and SH that interact with the 2D material represent BICs of a specially engineered Si metasurface \cite{Wang20242D}, as presented in Fig.~\ref{fig5}. More specifically, as illustrated in Fig.~\ref{fig5}(a), the nonlinear metasurface consists of a square array of cruciform-shaped Si nanoparticles, on top of which a monolayer of molybdenum disulfide (\ch{MoS2}) is deposited. It was demonstrated that by finely tuning the asymmetry parameter, $s = \Delta d/d$, of the silicon crosses, one can design a metasurface that possesses high-\textit{Q} optical quasi-BICs at the FF ($\omega$) and SH ($\Omega=2\omega$).

In this configuration, the maximum SHG in the \ch{MoS2} layer placed on top of silicon crosses was numerically calculated to be nearly three orders of magnitude larger than that corresponding to a suspended \ch{MoS2} layer, a maximum achieved for asymmetry parameter $s=0.63$; see also Fig.~\ref{fig5}(b). To gain deeper physical insights into this dual-BICs SHG enhancement mechanism, an eigenmode expansion method \cite{Koshelev2020} was applied to theoretically quantify the SHG intensity and then estimate the optical power generated at the SH. The conclusions of this theoretical analysis were validated by the results obtained from rigorous numerical simulations.

In addition to the characterization of the SHG in the nonlinear metasurface, the topological features of the two interacting quasi-BICs have also been investigated by calculating the far-field polarization and performing the Fourier analysis of the optical field corresponding to a symmetric metasurface. The topological charge of the FF BIC was determined to be equal to 1, as shown in the left panel of Fig.~\ref{fig5}(c). Importantly, considering the SH mode that governs the scattered nonlinear optical field, there exist 5 diffraction channels in the air region above the metasurface that can contribute to the decay of the energy in the optical resonance. Despite this, the radiative $Q$-factor of this optical mode is surprisingly large, namely $Q=5288$. As illustrated in the middle panel of Fig.~\ref{fig5}(c), this result was explained by the Fourier analysis of the mode at the SH, which proved that the energy leakage from the zeroth diffraction order vanishes. The underlying physics that explains this finding amounts to the symmetry incompatibility between the mode at the SH and the modes of the radiation continuum.

Regarding the topological properties of the BIC that exists at the SH, the far-field polarization of the nonradiative diffraction channel was shown to undergo a total phase change of $-2\pi$ along a circular path, indicating a topological charge of $q=-1$. The results are presented in the right panel of Fig.~\ref{fig5}(c). Moreover, it was found that the four first-order diffraction channels have trivial topological properties. Therefore, this SH mode can be viewed as a quasi-BIC mode, inasmuch as the zeroth diffraction order is completely decoupled from the radiation continuum whereas the higher-order diffraction channels are only weakly cupled to it.

\subsection{Nonlinear resonant chiral optical effects}
\label{3.3}
The chirality of an object refers to the property of its non-superimposable characteristics with respect to mirror symmetries. In the context of light-matter interaction, contrasting ways of interaction between light (intensity and phase) with right- and left-circularly polarized states and chiral structures bring about chiroptical effects, including optical activity, which amounts to the rotation of light polarization plane, and circular dichroism (CD), which describes the difference in absorption or transmission of circularly-polarized light upon interaction with chiral structures.

The concept of chirality is encountered in both linear and nonlinear optics. In the nonlinear optical regime, the strength of the nonlinear CD of harmonic generation is described by
\begin{equation}
    \eta_{CD}=\frac{I_{\mathrm{RCP}}^{n\omega}-I_{\mathrm{LCP}}^{n\omega}}{I_{\mathrm{RCP}}^{n\omega}+I_{\mathrm{LCP}}^{n\omega}},
    \label{Eq9}
\end{equation}
where $I_{\mathrm{RCP/LCP}}^{n\omega}$ is the $n$th-harmonic intensity under right-handed or left-handed circularly polarized (RCP or LCP) incidence. To date, photonic structures, including 3D metamaterials \cite{Lapine2014,Wang2016} and 2D metasurfaces \cite{Li2017,Valev2014,Chen2016,Valev2018,Koshelev2023}, have been used to demonstrate nonlinear chiroptical effects in various configurations. However, it is still rather challenging to obtain strong chiral optical effects facilitated by high-$Q$ optical resonances, thus restricting their deployment to a series of applications, such as sensing and chiral wave emission.
\begin{figure}[!t]
\centering
\includegraphics[width=\columnwidth]{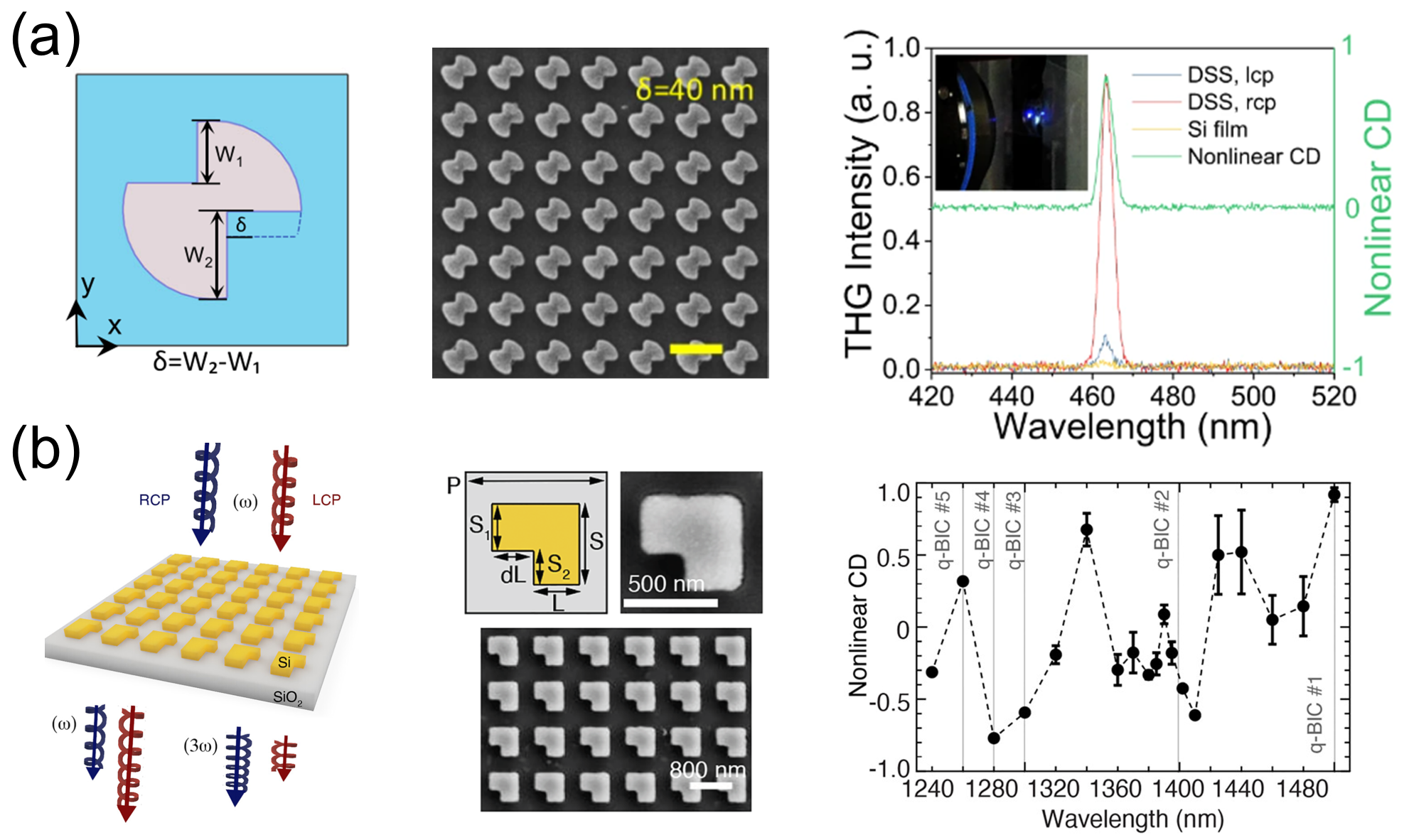}
\caption{Observation of nonlinear chiral effects from metasurfaces driven by BICs. (a) Near-unitary nonlinear circular dichroism from a double-sided scythe metasurface. Left panel, symmetry-broken unit cell designed by introducing $\delta=W_2 - W_1$. Middle panel, top view of the fabricated metasurface with optimized asymmetry parameter, $\delta=\SI{40}{\nm}$. Right panel, measured THG intensities under RCP/LCP incidences and the measured nonlinear CD spectra. Reproduced with permission from \cite{Shi2022}. (b) Resonant chiral effects in nonlinear dielectric metasurfaces. Left panel, schematic of an $L$-shaped Si nonlinear metasurface. Middle panel, unit cell design with asymmetry parameter $dL$ and the SEM image of the resonant metasurface. Right panel, measured TH CD spectrum marked with solid vertical lines for the five quasi-BIC resonances in case when $dL = \SI{300}{\nm}$. Reproduced with permission from \cite{Koshelev2023ACS}.}\label{fig6}
\end{figure}

In this context, BICs with large $Q$-factor provide a promising way to realize enhanced nonlinear chiral effects \cite{Gando2021,Shi2022,Koshelev2023ACS,Tonkaev2024,Liu2022,Cai2024,Duan2024}. Thus, in one of the first studies in which THG with large nonlinear CD was demonstrated, a photonic structure made of all-dielectric meta-atoms composed of two different Si blocks placed onto a silica substrate was employed \cite{Gando2021}. The meta-atoms were engineered so as to support quasi-BICs and to achieve both high frequency conversion efficiency and near-unity THG CD for RCP wave incidence in the near-IR frequency range. In the same study, it was also observed that by tuning the Si block length one can bring about both linear and nonlinear CD ($\eta_{CD}>0.8$) due to multimode interference. Subsequently, a dual-band chiral nonlinear metasurface was proposed to further enhance THG CD ($-0.94<\eta_{CD}<0.99$) with important applications to high-sensitivity chiral sensing being envisioned \cite{Liu2022}.

In a more recent study, a Si nonlinear resonant metasurface possessing intrinsic nonlinear (third and fifth orders) chirality was numerically demonstrated by splitting the topological charge of BICs \textit{via} in-plane and out-of-plane symmetry breaking \cite{Cai2024}, whereas a \ch{LiNbO3} nonlinear optical metasurface was proposed and numerically studied, whereby the SHG CD was driven by symmetry-protected BICs \cite{Duan2024}.

The experimental observation of nonlinear chiroptical response from chiral metasurfaces supporting BICs was not achieved until recently. Thus, it has been first realized experimentally in a planar nonlinear chiral metasurfaces made of Si and operating at optical frequencies \cite{Shi2022}. The double-sided scythe shaped unit cell configuration of the proposed Si metasurface is illustrated in the left panel of Fig.~\ref{fig6}(a), where an asymmetry parameter $\delta = W_2 - W_1$ is introduced to break the in-plane inversion symmetry. At the $\Gamma$-point, that is under normal incidence, such symmetry alteration ($\delta = \SI{40}{nm}$) converts the electric-dipole BIC state supporting a vortex polarization singularity into a right-handed circularly polarized eigenstate, thus giving rise to an intrinsically high-$Q$ chiral effect.

Both numerical and experimental investigations of the nonlinear CD have been performed in this study, the SEM image of the metasurface being shown in the middle panel of Fig.~\ref{fig6}(a). At the frequency of the quasi-BIC resonance, RCP pump beam induces significant enhancement of TH intensity, whereas THG under LCP excitation is negligible, leading to near unity nonlinear CD. The measured THG intensity and nonlinear CD are presented in the right panel of Fig.~\ref{fig6}(a), where the highest THG CD reaches 0.81 at a wavelength of $\sim$\SI{465}{\nm}. In addition, compared to a reference Si thin-film, the THG conversion efficiency is greatly enhanced by the strong light-matter interaction induced by the excitation of the quasi-BICs.

In a different work \cite{Koshelev2023ACS}, it was studied the relationship between the nonlinear CD enhancement from GMRs and quasi-BICs and the asymmetry of an $L$-shaped Si metasurface quantified by a parameter, $dL$, as per the left and middle panels of Fig.~\ref{fig6}(b). The sharp features of the resonances and the evolution of the $Q$-factor of the five quasi-BICs and one GMR were analyzed for frequencies ranging from \SIrange{1240}{1500}{\nm} and for values of $dL$ increasing from \SIrange{100}{300}{\nm}. Large THG efficiency and nonlinear CD were numerically and experimentally observed.

For metasurface samples with different asymmetries, the measured nonlinear THG CD varies from $0.918 \pm 0.049$ to $-0.771 \pm 0.004$. To illustrate this fact, the measured nonlinear THG CD in the case when $dL=\SI{300}{\nm}$ is presented in the right panel of Fig.~6(b). This points to a general strategy to change the nonlinear chiral response under RCP/LCP wave excitation while preserving large values of the TH intensity. Furthermore, it was reported that the nonreciprocity property of the nonlinear chirality, namely the difference in THG CD between right-handed and left-handed polarized pump sources, can be explained by the macroscopic symmetry of the $L$-shaped meta-atoms and the microscopic material nonlinearities.

More recently, bar-shaped meta-atoms made of nanostructured van der Waals materials were employed \cite{Tonkaev2024}, namely hexagonal-boron nitride (h-BN), to simultaneously achieve high nonlinear conversion efficiency and nonlinear CD. By designing the dielectric metasurface so as to support a quasi-BIC in the near-IR range, experimental measurements demonstrated a wide range of THG CD varying from -0.38 to +0.12 near the quasi-BIC resonance together with a huge THG enhancement of $\sim10^3$ as compared to a stand-alone h-BN thin film. It is expected that the exploration of nonlinear h-BN metasurfaces might lead to a new platform for chiral photonics based on nanostructured van der Waals materials.

\subsection{Complex nonlinear optical frequency mixing}\label{3.4}
The recent advances in metasurface-mediated nonlinear harmonic generation and harmonic chiroptical effects arising from the local field enhancement and confinement in conjunction with the symmetry-governed vortex center features of optical BICs have been reviewed in the preceding subsections. In addition to these nonlinear optical processes, which generate photons with frequencies at integer multiples of the frequency of the incoming photons, more complex nonlinear frequency mixing processes, including degenerate FWM \cite{More2024,Xu2022,More2021,Liu2023}, SPDC \cite{San2022,Parry2021,Mazza2022,Zhang2022SPDC,Son2023,Liu2024}, SFG \cite{Camacho2022,Cai2023}, and OR \cite{Hu2022}, have also been experimentally demonstrated. These nonlinear optical interactions greatly broaden the application perspectives of BIC-governed nonlinear metasurfaces in optics and quantum technologies.

\begin{figure}[!t]
\centering
\includegraphics[width=\columnwidth]{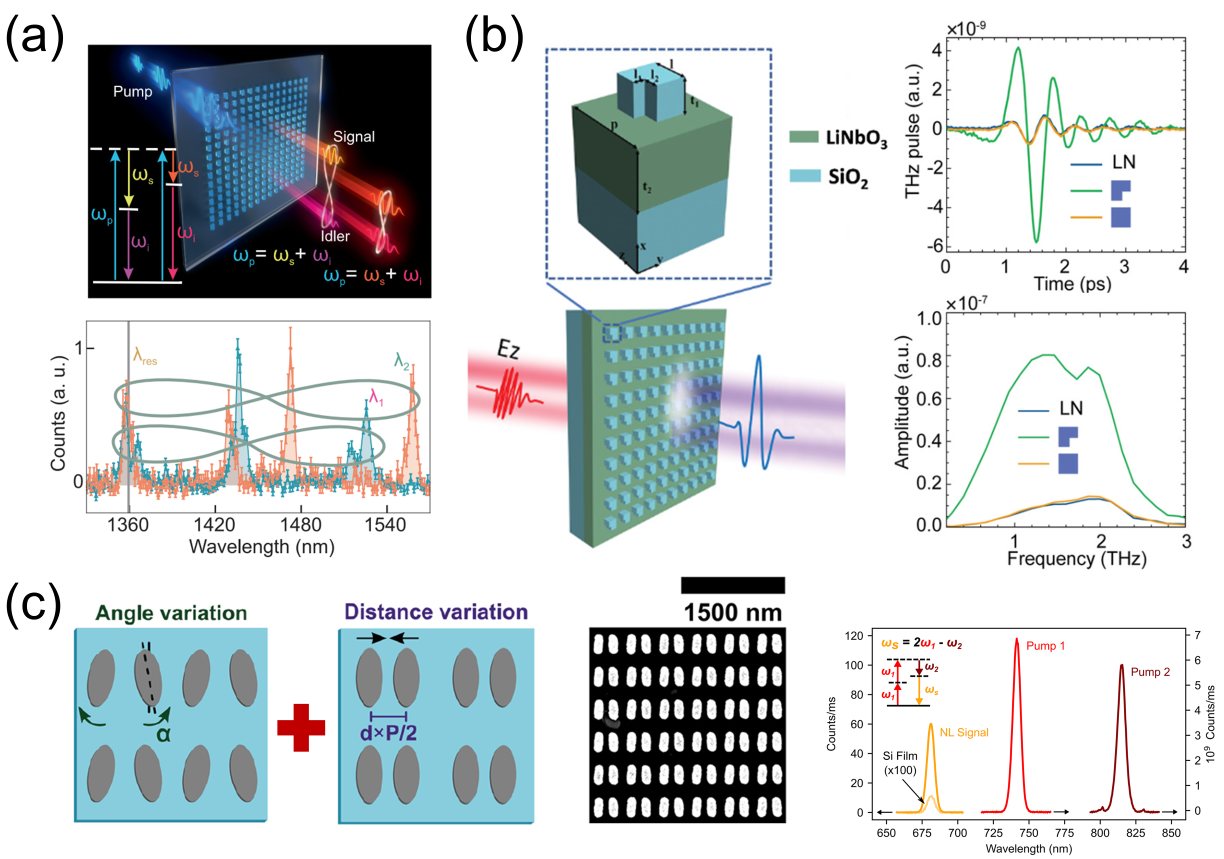}
\caption{Application of BICs to intricate frequency mixing processes supported by metasurfaces. (a) Complex quantum states generation from SPDC. Top panel, schematic of entangled photon generation in symmetry-broken GaAs metasurfaces. Bottom panel, complex cluster quantum states generation under multiple pump beams by utilizing electric-dipole and magnetic-dipole BICs. Reproduced with permission from \cite{San2022}. (b) Broadband terahertz generation from \ch{LiNbO3} thin-film \textit{via} OR. Left panel, schematic illustration of the metasurface design. Top right panel, time-resolved terahertz pulses of bare thin-film, $L$-shaped metasurface, and square-shaped metasurface. Bottom right panel, Fourier transform spectra of the terahertz emission. Reproduced with permission from \cite{Hu2022}. (c) Silicon-on-silica metasurface for degenerate four-wave mixing. Left panel, symmetry distortion \textit{via} angle tilt and intra-cell distance. Middle panel, SEM image of a fabricated metasurface. Right panel, spectra of pump lasers (Pump 1 and Pump 2) and generated nonlinear signal (NL Signal). Reproduced with permission from \cite{More2024}.}\label{fig7}
\end{figure}

Regarding SPDC in BIC-governed metasurfaces, it was recently proposed to employ a certain nonlinear BIC-resonant metasurfaces made of GaAs for generating complex quantum states, namely to generate from a pump photon ($\omega_p$) of higher frequency a pair of signal ($\omega_s$) and idler ($\omega_i$) photons of lower frequencies \cite{San2022}, as illustrated in the top panel of Fig.~\ref{fig7}(a). The optical energy conservation in this parametric process gives $\omega_p = \omega_s + \omega_i$. Utilizing SPDC arising from metasurfaces with subwavelength thickness, the resulting optical quantum states can be engineered without the constraint imposed by the momentum conservation law in conventional nonlinear optical waveguides and crystals.

The BIC-empowered high-$Q$ optical resonances of nonlinear metasurfaces have the potential to greatly boost the spontaneous emission of photons through the enhancement of the vacuum field fluctuations. In this context, it was experimentally demonstrated that, by breaking the rotational symmetries $C_2$ and $C_4$ of its constituent meta-atoms, it is possible to design a nonlinear metasurface that supports an electric-dipole quasi-BIC at \SI{1446.9}{\nm} with $Q$-factor, $Q\sim 330$, and a magnetic-dipole quasi-BIC at \SI{1511.8}{\nm} with $Q\sim 1000$ \cite{San2022}. It was demonstrated that, by taking advantage of the sharp resonance features of quasi-BICs, one can generate degenerate ($\omega_s = \omega_i$) and nondegenerate ($\omega_s \neq \omega_i$) entangled photon pairs characterized by a broad tunable spectral range (more than \SI{100}{\nm}) without any decrease of the frequency conversion efficiency. The enhancement of generation of photon pairs reaches at least three orders of magnitude, as compared to the case of a GaAs thin-film.

In addition, multifrequency quantum states, such as cluster states, can be generated from a single or several quasi-BIC resonances when pumped with continuous-wave laser beams of multiple wavelengths \cite{San2022}, as shown in the bottom panel of Fig.~\ref{fig7}(a). In a different work, the SPDC process in \ch{LiNbO3} metasurfaces that support BICs has been used experimentally to enhance spatially entangled photon-pair rate by 450 times in the telecommunication band \cite{Zhang2022SPDC}, whereas bi-directional emission of entangled photons upon SPDC interaction in a GaP metasurface supporting BICs  have also been observed \cite{Son2023}. Besides this experimental work, advances in theoretical understanding of the physics of enhanced photon pair generation facilitated by BICs have also been reported \cite{Parry2021,Mazza2022,Liu2024}. From a practical perspective, these investigations reveal novel opportunities of using BIC-based metasurfaces as a versatile platform for quantum optics and signal processing.

All-dielectric metasurfaces supporting quasi-BICs have also been used to generate bright and broad THz emission by employing the second-order nonlinearity of \ch{LiNbO3} films under near-IR fs laser pulse excitation. This represents a simple yet novel path towards the realization of broadband THz emitters \cite{Hu2022}. As presented in the left panel of Fig.~\ref{fig7}(b), the metasurface consists of a cubic-lattice patterned $L-$shaped silicon dioxide (\ch{SiO2}) and a $x$-cut \ch{LiNbO3} thin layer, placed onto a \ch{SiO2} substrate. Such meta-atom design avoids certain complications and challenges, including large structure depths and steep sidewalls in the etching process of \ch{LiNbO3} nanostructures. Through careful tuning of the metasurface geometry, strong coupling between a magnetic-dipole BIC and an electric-dipole Mie resonance is achieved, leading to the localization of the optical field in the high-index, nonlinear \ch{LiNbO3} film instead of the patterned \ch{SiO2} layer.

The experimental results reported in this study are illustrated in the right panel of Fig.~\ref{fig7}(b), including both measured time-domain THz pulses and their Fourier transformed spectra. Driven by the ultrashort laser pulse with width of \SI{70}{\fs}, and operating at \SI{800}{\nm} \textit{via} OR, it was observed a 17-fold enhancement of THz amplitude at \SI{0.7}{\THz} as compared to the square-shaped metasurface and bare \ch{LiNbO3} thin-film. Importantly, the frequency range of the enhanced THz emission effect spans a wide spectral domain, from \SIrange{0.1}{4.5}{\THz}, implying an efficient mechanism for an optical source for THz generation.

In the context of quadratic nonlinear frequency mixing involving two input beams at different frequencies, SFG is one of the most thoroughly studied optical process. For example, it has been experimentally demonstrated that nonlinear SFG in multiresonant GaP metasurfaces based on doubly resonant waveguide-type BICs can be efficiently generated \cite{Camacho2022}. The experiment involved two near-IR input beams, a pump (signal) beam at \SI{875}{\nm} (\SI{1545}{\nm}) with  intensity of \SI{137}{\mega\watt\per\square\cm} (\SI{12.5}{\mega\watt\per\square\cm}), and was observed strong SFG in the visible spectrum, at \SI{558}{\nm}, realizing a high normalized up-conversion efficiency of \SI{2.5e-4}{\per\watt}. Interestingly, the two incoming optical beams possessed a nonparallel polarization state, which allowed independent control of the nonlinear light generation in different diffraction channels. Moreover, the high-\textit{Q} feature of BIC-induced resonances was also used to study SFG from etchless \ch{LiNbO3} thin layers, as theoretically demonstrated in Ref.~\cite{Cai2023}. This approach circumvents the challenges encountered in the manufacturing process of etching \ch{LiNbO3} thin-films.

Four-wave mixing, a third-order nonlinear optical process, involves the interaction of two or three photons with different frequencies that results in the generation of a photon with a new frequency. For the three-photon interaction, called degenerate FWM, the pump consists of only two frequency components, $\omega_1$ and $\omega_2$, the resulting photon being emitted at the frequency imposed by the energy conservation law, $\omega_s = 2\omega_1 - \omega_2$, where $\omega_s$ is the generated nonlinear signal.

Dielectric metasurfaces possessing Mie resonances and BICs offer unique possibilities for the realization of multiple high-$Q$ optical resonances while relaxing the phase-matching conditions. This makes them a versatile platform to implement FWM and other nonlinear optical interactions. To this end, a multi-resonant Si dimer-hole metasurface was designed, which simultaneously supports a toroidal dipole BIC in the short-wave IR spectrum and a Mie resonance in the near-IR region \cite{Xu2022}. Using degenerate FWM, visible light radiation was observed in this experimental investigation. With low pump peak intensities of \SI{0.33}{\giga\watt\per\square\cm} and \SI{0.17}{\giga\watt\per\square\cm} at the two resonances, a large nonlinear conversion efficiency of \num{0.76e-6} was measured.

Metasurfaces that utilize only multiple BICs for FWM have also been studied. In particular, numerical investigations have shown that GaP nanostructures can be engineered so as to enhance degenerate FWM \cite{More2021}, whereas in a different work Si nanodisk dimer metasurfaces were used for both FWM and degenerate FWM \cite{Liu2023}. Experimental demonstration of FWM in purely BIC-assisted multi-resonant metasurfaces has recently been reported, where a silicon-on-silica metasurface supporting four BICs in the near-infrared region has been introduced \cite{More2024}.

Combining both angle tilt and intra-cell distance in the meta-atom geometry design methodology, as illustrated in the left panel of Fig.~\ref{fig7}(c), one obtains four quasi-BICs induced by symmetry breaking that are characterized by narrow-band spectral features, with the $Q$-factor ranging from \numrange{50}{200}. The measured normalized FWM efficiency, $P_s/(P_1^2 P_2)$, was reported to be \SI{1.1 \pm 0.2}{\per\square\watt\percent} under average pump laser power of \SI{1}{\mW}. Moreover, compared to an unstructured Si thin-film with the same thickness as that of the metasurface, the generated FWM nonlinear signal was enhanced by 600 times, which was an order of magnitude larger than the value reported in the work that used both quasi-BICs and Mie resonances \cite{Xu2022}, as per the right panel of Fig.~\ref{fig7}(c).

\section{Conclusion and Outlook}

\begin{table}[!t]
\caption{Summary of nonlinear optical processes in BIC-based dielectric metasurfaces.}
\label{table1}
\centering
\renewcommand{\arraystretch}{1.5}
\begin{adjustbox}{width=\textwidth}
\begin{tabular}{c c c c c c c }
\hline
\textbf{Type} & \textbf{Function} & \textbf{BIC type} & \textbf{$Q$-factor} & \makecell{\textbf{Nonlinear} \\ \textbf{efficiency}} & \textbf{Material} & \textbf{Ref.} \\
\hline

\makecell{Harmonic \\ generation} & SHG & \makecell{Symmetry- \\ protected BIC} & $\sim500$ & $\sim6\times10^{-6}$ & GaAs & \cite{Vabi2018} \\

 & SHG & \makecell{Symmetry- \\ protected BIC} & 651 & $\sim6.8\times10^{-7}$ & Si & \cite{Fang2022} \\

 & THG & \makecell{Symmetry- \\ protected BIC} & $\sim88$ & $\sim10^{-6}$ & a-Si & \cite{Koshelev2019THG} \\

& THG & \makecell{Symmetry- \\ protected BIC} & $\sim600$ & $1.8\times10^{-6}$ & a-Si & \cite{Yang2022} \\

& THG & \makecell{Symmetry- \\ protected BIC} & 18511 & $9.1\times10^{-7}$ & c-Si & \cite{Liu2019} \\

& THG & \makecell{Accidental \\ BIC} & 206 & $1.13\times10^{-5}$ & a-Si & \cite{Tang2024} \\

& \makecell{HHG \\ ($5\omega, 7\omega,$ \\ $9\omega, 11\omega$)} & \makecell{Symmetry- \\ protected BIC} & 50 - 100 & \makecell{$\sim600^*$ ($5\omega$) \\ $\sim18^*$ ($7\omega$) \\ $\sim3^*$ ($9\omega$) \\ N/A ($11\omega$)} & a-Si & \cite{Zogra2022} \\

\hline

\makecell{Harmonic \\ generation \\ (2D materials)} &  SHG & \makecell{Symmetry- \\ protected BIC} & $\sim35$ & $1140^*$ & \ch{WS2} monolayer & \cite{Bernhar2020} \\

 & SHG & \makecell{Symmetry- \\ protected BIC} & 8911 & $\sim9400^*$ & GaSe flake & \cite{Liu20212D} \\

 & SHG & \makecell{Symmetry- \\ protected BIC} & $\sim75$ & $\sim170^*$ & \ch{WSe2} monolayer & \cite{Ren2024} \\

\hline

\makecell{Nonlinear \\ chiroptical \\ effects} & THG CD & \makecell{Symmetry- \\ protected BIC} & $\sim100$ & [0, 0.81]$^{\triangledown}$ & a-Si & \cite{Shi2022} \\

 & THG CD & \makecell{Symmetry- \\ protected BIC} & 50 - 300 & [-0.77, 0.92]$^{\triangledown}$ & a-Si & \cite{Koshelev2023ACS} \\

 & THG CD & \makecell{Symmetry- \\ protected BIC} & $\sim100$ & [-0.38, 0.12]$^{\triangledown}$ & hBN & \cite{Tonkaev2024} \\

\hline

\makecell{Complex \\ frequency \\ mixing} & OR & \makecell{Symmetry- \\ protected BIC}  & 453 & 17$^*$ & \ch{LiNbO3} & \cite{Hu2022} \\

 & SFG & \makecell{Waveguide- \\ type BIC} & \makecell{$\sim173$ ($\omega_p$) \\ $\sim103$ ($\omega_s$)} & $6.4\times10^{-6}$  & GaP & \cite{Camacho2022} \\

 & SPDC & \makecell{Symmetry- \\ protected BIC} & \makecell{$\sim330$ ($\omega_s$) \\ $\sim1000$ ($\omega_i$)} & $\geq10^{3*}$ & GaAs & \cite{San2022} \\

 & FWM & \makecell{Toroidal \\ dioole BIC} & $\sim180$ ($\omega_s$) & $0.76\times10^{-6}$ & a-Si & \cite{Xu2022} \\

 & FWM & \makecell{Symmetry- \\ protected BIC} & \makecell{50 - 200 \\ ($\omega_{p,1}$, $\omega_{p,2}$)} & $\sim1.1\times10^{-8}$ & a-Si & \cite{More2024} \\

 & \makecell{Self-action \\ effect} & \makecell{Symmetry- \\ protected BIC} & $\sim900$ & N/A & Si & \cite{Sinev2021} \\
\hline
\end{tabular}
\end{adjustbox}
\begin{tablenotes}
\item{ $\omega_p$, $\omega_s$, $\omega_i$: pump frequency, signal frequency, and idler frequency, respectively.}
\item{ $*$: nonlinear enhancement factor is used where nonlinear conversion efficiency is not available.}
\item{ $\triangledown$: nonlinear CD is used where nonlinear conversion efficiency is not available.}
\end{tablenotes}
\end{table}

In this review, we have outlined and discussed the physics and some of the unique properties of a special type of optical resonances, called bound-states in the continuum, with a specific focus being placed on their use in nonlinear optics. Our discussion covered both fundamental aspects of the role played by bound-states in the continuum in nonlinear optics, as well as their potential influence on the development of future highly impactful nonlinear optical applications. We have tried to keep at the center of our presentation those properties that make the bound-state in the continuum so distinct from most other optical resonances, namely their topological properties, optical field localization, and practically infinite life-time, and highlight those applications derived from these unique properties. In particular, we have highlighted the role that bound-states in the continuum in metasurfaces can play in enhancing nonlinear optical processes that are at the heart of key technological applications. To summarize, we list in Table~\ref{table1} and compare research results pertaining to nonlinear optical metasurfaces empowered by different types of BICs.

Regarding BIC-assisted nonlinear light-matter interactions in metasurfaces, we first surveyed recent advances in harmonic generation processes with orders ranging from second to eleventh. We illustrated the significant role of critical coupling conditions in maximizing the nonlinear signal generation in metasurfaces with broken symmetry. We discussed the integration of 2D layered materials with BIC-driven dielectric metasurfaces, which revealed a versatile approach for boosting the effective second-order nonlinear susceptibility of 2D materials. Then, we presented several studies of the generation of high harmonics assisted by the excitation of quasi-BICs in nonlinear metasurfaces and highlighted the relationship between this process and nonperturbative nonlinear optical processes at the nanoscale. Moreover, we also summarized recent demonstrations of enhanced even-order nonlinear optical processes in centrosymmetric media by utilizing field confinement effects arising from excitation of BICs in nonlinear metasurfaces.

We also presented and discussed a mechanism commonly used to enhance nonlinear optical interactions in metasurfaces and the efficiency of nonlinear frequency conversion, namely the excitation of multiple optical resonances. Among others, we highlighted a recent study in which it was demonstrated the application of this multi-resonance mechanism to a nonlinear metasurface integrated with 2D TMD materials, whereby a  pair of BICs (a symmetry-protected BIC at the fundamental frequency and a quasi-BIC at the second harmonic) induced strong enhancement of the intensity of SHG.

Moreover, we discussed the recent progress in the field of nonlinear resonant chiral optical effects driven by dielectric metasurfaces based on BICs. Taking advantage of the optical field confinement and asymmetry-dependent out-coupling intensity of BICs, intense and tunable nonlinear chiroptical response as well as large nonlinear conversion efficiency has been demonstrated using symmetry-broken meta-atoms. In addition, intrinsic circular eigen-polarization stemming from the nature of vortex polarization singularity of BICs has been employed to generate nonlinear CD in chiral metasurfaces. The application of these findings to nonlinear chiroptical nanodevices has also been discussed.

As the last reviewed topic, we surveyed the application of BIC-governed dielectric metasurfaces to complex frequency mixing processes. In this context, we illustrated the use of SPDC in BIC metasurfaces for generating complex quantum states and discussed specific applications to quantum optics and information processing. The demonstration of another practical application of BIC-based dielectric metasurfaces to broadband, highly efficient, and compact THz sources was also discussed. Moreover, we presented several studies whereby SFG and FWM have been numerically and experimentally realized in nonlinear metasurfaces consisting of dielectric materials, in which the high-$Q$ BICs greatly enhance the nonlinear frequency efficiency.

The field of nonlinear optics of bound-states in the continuum is still in its infancy, so that one expects to witness in the future a plethora of new and exciting developments in this area, both at the fundamental science level and in relation with the emergence of new devices with new or improved functionalities. For example, it is well-known that nonlinear optical interactions are strongly influenced by the requirement that certain physical quantities are conserved during the nonlinear interactions, constraints described, e.g., by the Manley-Rowe relations. Regarding this matter, a clear understanding of how the topological properties of bound-states in the continuum affect these constraints is still missing. Moreover, at a more applied level, it is now well understood the impact that bound-states in the continuum can have on the development of future ultra-compact lasers, yet very little is known about how nonlinear optical effects can affect the corresponding lasing modes. Looking for the answers to these and many other exciting questions will ensure that the field of nonlinear optics of bound-states in the continuum will continue to grow at an accelerating pace.

\section*{Declaration of competing interest}
The authors declare that they have no known competing financial interests or personal relationships that could have appeared
to influence the work reported in this paper.




\end{document}